# An ANN Approach in Predicting Solar and Geophysical Indices from Ionospheric TEC Over Indore

Sumanjit Chakraborty[✉] and Abhirup Datta

Discipline of Astronomy, Astrophysics and Space Engineering, Indian Institute of Technolgy Indore, Simrol, Indore 453552, India
{phd1601121006,abhirup.datta}@iiti.ac.in

**Abstract.** In this paper, preliminary results from the artificial neural network (ANN)-based model developed at IIT Indore have been presented. One year hourly total electron content (TEC) database has been created from the International Reference Ionosphere (IRI)—2016 model. For the first time, a reverse problem has been addressed, wherein the training has been performed for predicting the three indices: 13-month running sunspot number, ionospheric index and daily solar radio flux also called targets to the network when hourly TEC values are the inputs. The root mean square errors (RMSEs) of these targets have been compared and minimized after several training of the dataset using different sets of combinations. Unknown data fed to the network yielded 0.99%, 3.12% and 0.90% errors for Rz12, IG12 and F10.7 radio flux, respectively, thus signifying ~97% prediction accuracy of the model.

**Keyword:** Solar indices · Geophysical indices · Ionospheric TEC · IRI · Machine learning · ANN

## 1 Introduction

The Earth's ionosphere is formed as a consequence of ionization by solar radiation. It extends from about 60–1000 km above the Earth's surface. Since nearly 70% of the global ionization is concentrated in and around ±15° magnetic latitude crests due to the equatorial ionization anomaly (EIA) [1], it becomes essential to characterize dynamism of the variable ionosphere over this region where sharp latitudinal gradient exists. The total electron content (TEC), which is the columnar number density of electrons expressed in TEC units (1 TEC unit = $10^{16}$ electrons/m$^2$), is a fundamental observable parameter [6] which plays an important role in characterization of the ionosphere. The location chosen for the analysis, Indore (22.52° N, 75.92° E geographic) falls near the northern crest of EIA and as a result is a suitable location to study the variability of the ionosphere. Complexity of the spatial and temporal variations of the ionosphere makes it difficult to characterize or model the ionosphere and accurately forecast its impact on the global navigation satellite system (GNSS) signals. Therefore, a requirement arises for the development of an artificial neural network (ANN)-based model that would be able to predict ionospheric behavior over the regions where physical data is unavailable.

ANNs, inspired from the neural networks which constitute animal brains, have collection of connected nodes known as artificial neurons. Each connection similar to the synapses in a biological brain can transmit signal from one artificial neuron to another [8]. These connections between neurons are called edges. Artificial neurons and these edges have a weight that self adjust as learning proceeds. This weightage may increase or decrease the strength of the signal at a connection. The neurons have a threshold above which the signal is sent and are aggregated into different layers which perform various transformations on their inputs. The signals travel from the input (first layer), traverses multiple layers (hidden layers) and arrives to the output (last layer) [4]. The ANN discussed in this paper has been developed by preparing dataset from the empirical International Reference Ionosphere (IRI) model. The sources of data to this model are the dense global network of ionosondes, incoherent scatter radars, and Alouette topside sounders in situ instruments on board satellites. Inputs to this model are the date, latitude, longitude and topside electron boundary while the outputs are electron temperature and density, ion temperature and composition and the TEC from 50 to 2000 km altitude range [7].

Studies have been made by several researchers [2, 9, 10] to predict TEC model of the ionosphere by auto-regressive method. Studies have also been made [3, 5] in development of ANN-based TEC model where the solar and geophysical indices are fed as inputs to obtain predicted TEC as output, but for the first time, to the best of our knowledge, the TEC data has been fed as network inputs to obtain the solar and geophysical indices. This work thus addresses a reverse problem, wherein by having the knowledge of TEC variation, one could be able to infer about the indices that would be vital in understanding space weather and ensure flawless service to GNSS users.

## 2 Methodology

A feed forward network has been used where the signal gets propagated from the input layer to the hidden layer and then to the output layer. The present model is generated by using a single hidden layer of 25 neurons. The model inputs are the hourly vertical TEC values over Indore obtained from the IRI-2016 Web model for the entire year of 2017 which had been in the declining phase of solar cycle 24. The targets set to this model are the 13-month running mean of sunspot number (Rz12), the ionospheric index (IG12) and the F10.7 daily radio flux (sfu). Connections between the nodes are such that they represent the feeding of the output from one node to the other, multiplied by a weight. The weights given to the hidden layer are appropriately modified to obtain relatively less prediction error between the targets and predicted indices. The optimized architecture for the network is obtained by trial and error while the biases and weights are adjusted according to the Levenberg–Marquardt algorithm. The architecture of an ANN with an input layer, a single hidden layer and an output layer is depicted in Fig. 1.

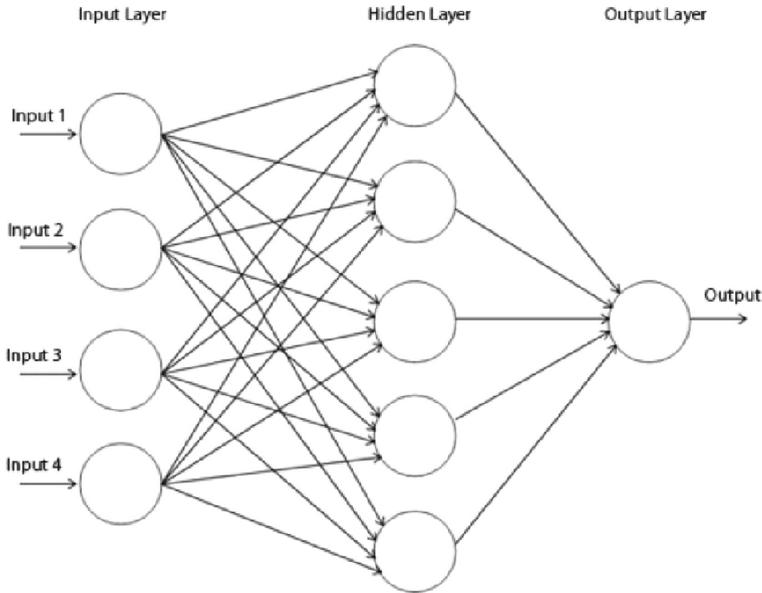

**Fig. 1** Typical ANN structure showing the input, hidden and output layers (https://www.analyticsindiamag.com/artificial-neural-networks-101/)

## 3 Results

For training the neural network, 300 days were randomly selected from a one year database. The splitting of data for training, validation and testing was randomly selected as 70%, 15% and 15%, respectively. The activation function found to be suitable was tan sigmoid given by:

$$\tan \text{sig}(x) = \frac{1 - e^{-2x}}{1 + e^{-2x}} \quad (1)$$

This activation function is used to introduce nonlinearity to the network. This helps the network to understand the complexity and give accurate results. The error function at the end of one feed forward process to check training performance was the mean squared error (MSE) given by the mean of the squared of the error defined as the difference between the predictions and the targets. The idea is to minimize this error function by assigning suitable weights and biases at every step. The network was then trained several times by changing the number of neurons until the cost function was minimum that would perform well when subjected to unknown data. The remaining 65 days that is unknown to the trained network was used in order to check the model performance. Figure 2 shows the scatter plot of predictions and target with the 1:1 red line signifying an accuracy of 100%. The normalized RMSE values obtained are 0.0099, 0.0312 and 0.0090 translating to percentage errors of 0.99%, 3.12% and 0.90% for the indices Rz12, IG12 and F10.7 radio flux, respectively. These RMSE values are computed by using:

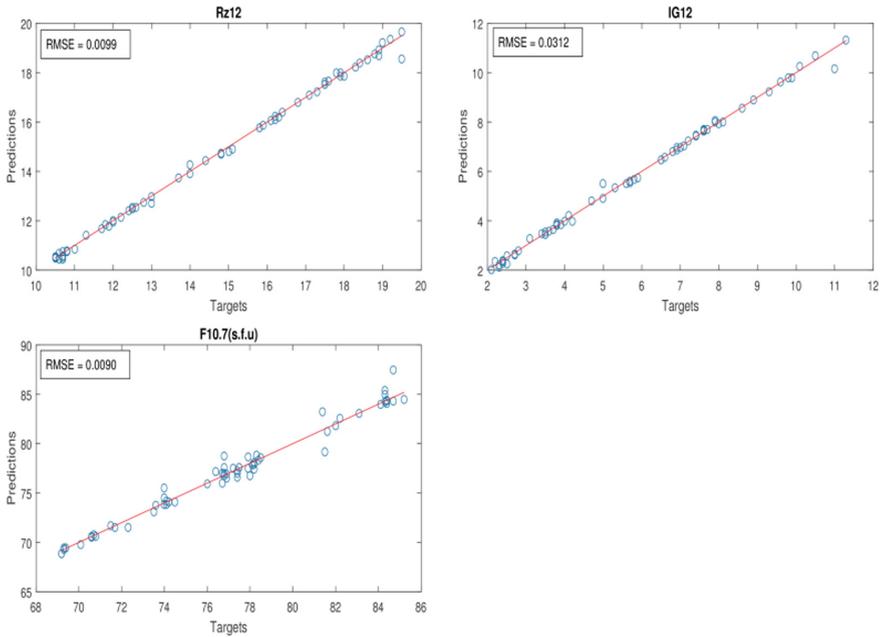

**Fig. 2.** Scatter plot showing the true values (targets) and the outputs (predictions) values from the neural network with one hidden layer of 25 neurons

$$\sqrt{\frac{1}{N}\sum\left(\frac{\text{targets}-\text{predictions}}{\text{targets}}\right)^2} \qquad (2)$$

## 4 Conclusions

The paper presents initial results from the ANN model developed over Indore, which is near to the anomaly crest, using the IRI model derived database to predict solar and geophysical indices. A single hidden layer has been used and a number of neurons have been varied to obtain the optimized results of this model. The normalized RMSE gave about 0.99, 0.90 and 3.12% errors in Rz12, F10.7 and IG12 index, respectively. Thus, ~97% accuracy has been achieved when unknown data was fed to the trained network. A reverse problem approach is addressed, wherein with the knowledge of TEC, prediction of various indices can be obtained even if real-time indices are not available. Path forward to this work would be to use this approach in training the network with real data and compare with the presented work for validation. This model could help in understanding the variable ionosphere where real data is unavailable.